# ARTICLE

## Protein-based microsphere biolasers fabricated by dehydration

Toan Van Nguyen,[a,b] Nhat Van Pham,[c] Hanh Hong Mai,[b] Dung Chi Duong,[d] Hai Hoang Le,[d] Riccardo Sapienza[e] and Van-Duong Ta*[d]

Biolasers made of biological materials have attracted a great of research attention due to their biocompatibility and biodegradability, which have the potential for biosensors and biointegration. However, the current fabrication method of biolasers suffers several limitations such as complicated processes, time-consuming and environmental unfriendly. In this work, a novel approach with green processes for fabricating solid-state microspheres biolasers is demonstrated. By using dehydration via a modified Microglassification™ technology, dye-doped bovine serum albumin (BSA) droplets can quickly (less than 10 minutes) and easily turn into solid microspheres with the diameter ranging from 10-150 μm. The size of the microspheres can be effectively controlled by changing either the concentration of BSA solution or the diameter of the initial droplets. Fabricated microspheres can act as efficient microlasers under optical pulse excitation. Lasing threshold of 7.8 μJ mm⁻² and quality (Q) factor of about 1700 to 3100 are obtained. The size-dependence of lasing characteristics have been investigated and the results show a good agreement with whispering gallery mode (WGM) theory. Our finding contributes an effective technique for the fabrication of high Q factor microlasers that may be potential for applications in biological and chemical sensors.

## 1. Introduction

Biolasers - laser sources composed of materials of biological origin - have attracted a great deal of interest due to their potential applications in bio-integration and biosensing.[1-4] Various biological materials including poly lactic-co-glycolicacid, starch, protein, pectin, cellulose, curcumin have been explored for laser microcavities.[5-9] Among these, bovine serum albumin (BSA) is considered to be an excellent biomaterial because of its biocompatibility and the ability to be transported in the human body.[10, 11] Along with developments on new materials, new lasing architectures, as for example random lasing,[12, 13] Fabry-Perot,[14, 15] distributed feedback,[16, 17] and whispering gallery mode (WGM) cavities[18-21] have been developed for biolasers. Particularly, microspheres biolasers are very interesting owing to their simple fabrication, high quality (Q) factor, low lasing threshold and promising for ultra-sensitive biosensors with dimension down to the intracellular level.[22, 23]

Microsphere biolasers can be fabricated using several techniques such as freeze-drying vacuum,[7] slow solvent evaporation,[9] formation oil droplets in cells,[19] emulsion and dehydration of droplets in polydimethylsiloxane (PDMS).[8] Even though the available methods are effective for making microsphere biolasers, there are still several limitations including complicated multiple fabrication processes,[7, 19] environmental unfriendly (using solvent)[8] and time-consuming (up to more than 12 hours)[7]. As a result, a novel technique that allows obtaining microsphere biolasers in a short time (less than 10 minutes) and with green processing is important for the development of biolasers.

In biology, dehydration is a common technique used for producing solid formulation of biologics. Particularly, a recently developed process, the so-called Microglassification™, is an effective method for fabricating solid microspheres rapidly and controllably from a protein solution.[24] In this technique, protein droplets can turn into solid spheres quickly by the diffusion of water molecules outside the droplet to the appropriated outer environment as decanol, pentanol until the dynamic equilibrium is established.[25] Surprisingly, the use of this approach for the fabrication of microsphere biolasers has not been explored so far. In this paper, we demonstrate the fabrication and optical characterizations of high-quality microsphere biolasers based on a modified Microglassification™ technology.

## 2. Experimental

**Preparation of dye-doped BSA solutions**

Bovine serum albumin (BSA, ≥ 98% purity), Rhodamine B (RhB, ≥ 95% dye), and 1-decanol (the so-called decanol for short, ≥ 99% purity) were purchased from Sigma-Aldrich. First, BSA solutions with four different concentrations were obtained by dissolving 1 g of BSA in 1, 4, 7 and 9 mL deionized water. Subsequently, 1 mL of RhB 1 wt% aqueous solutions were added to the four BSA solutions above. As a result, RhB doped BSA solutions with BSA concentrations of 500, 200, 125 and 100 mg/mL were obtained. The dry ratio of RhB and BSA in the four solutions is the same, which is 99 wt% of BSA and 1 wt% of RhB.

**Fabrication of microsphere biolasers**

First, a micropipette was used to create droplets with different sizes in decanol. It is noted that the decanol was stored in a Teflon

a. Department of Physics, Le Quy Don Technical University, Hanoi 100000, Vietnam.
b. Department of Quantum Optics, Faculty of Physics, VNU University of Science, Hanoi 100000, Vietnam.
c. Department of Advanced Material Science and Nanotechnology, University of Science and Technology Hanoi VAST, Hanoi 100000, Vietnam.
d. Department of Optical Devices, Le Quy Don Technical University, Hanoi 100000, Vietnam.
e. The Blackett Laboratory, Department of Physics, Imperial College London, London, SW7 2AZ, UK
*E-mail of corresponding author: duong.ta@mta.edu.vn
†Electronic Supplementary Information (ESI) available: Movie 1 demonstrates the formation of a solid microsphere biolaser from a droplet via dehydration. See DOI: 10.1039/x0xx00000x

container (instead of ordinary glass beaker) due to its hydrophobic property which allows obtaining perfect solid microspheres. The droplets were kept in decanol until the dehydration process is completed (about 10 minutes). Then, the obtained solid microspheres (which settle down at the bottom of the container) were taken out and subsequently heated at 100° C for 5 minutes (to completely remove the decanol). Finally, the microspheres were left to cool down at room temperature in ambient conditions.

**Optical characterizations**

An optical microscope with a magnification of 10× and 40× combined with a camera was used to monitor the formation of microspheres. During the dehydration process, the camera captured the droplets continuously after equal intervals of 6 s. Droplet sizes were determined by using MATLAB software. In addition, the surface morphology of microspheres was studied by SEM (TM4000plus-HITACHI). The microspheres were coated with a thin gold layer of about 10 nm thickness (by sputtering) before SEM analysis.

A micro-photoluminescence (μ-PL) setup was used to study the obtained microspheres. The pumping source is a Nd:YAG nanosecond pulse laser (Litron Lasers) with a wavelength of 532 nm, a repetition rate of 10 Hz and a pulse duration of 4-7 ns. The microspheres were excited by a focus laser beam with a spot size of ~ 350 μm in diameter. Emission from the microspheres was then collected by a 10× objective and subsequently delivered to an AvaSpec-2048L (Avantes) for spectral recording. The spectral resolution is ~ 0.2 nm. All optical characterizations were carried out in the air, at room temperature, and in ambient conditions.

## 3. Results and discussion

Fig. 1 shows the dehydration process starting as soon as a liquid droplet is created inside decanol. The aqueous droplet contains BSA and Rhodamine (RhB) dye molecules. The solubility of decanol in water is very low which allows us neglecting decanol diffusion into micro-droplets.[26] On the other hand, the solubility of water into decanol is high enough to ensure that the amount of water in the droplet is much lower than the dissolution capability of the surrounding medium. Based on this effect, Rickard *et al.* and Aniket *et al.* have successfully studied the dehydration of BSA and Lysozyme using pipette technique to obtain the rigid sphere of protein.[24, 25] In these studies, the system can be seen as the one-way diffusion of water into decanol and the diffusion of protein to decanol and decanol to water can be ignored. This process is expected to be similar to the dehydration of our droplets. With regards to RhB molecules, they are dissolved in decanol however the portion of RhB molecules leaking to decanol is negligible thanks to the hydrophobic and electrostatic interactions binding of the RhB molecules to BSA molecules.[27] The dehydration process is completed when the thermodynamic equilibrium is reached. Whereas, nearly all water molecules are removed from the droplets and solid-state microspheres of dye-doped BSA are formed. [25, 28]

Fig. 2 presents the optical microscope and SEM images of the dye-doped BSA microspheres. Their size can vary flexibly from 10 to 150 μm depending on the experimental conditions. In Fig. 2a, the microspheres are uniformly dark red. In addition, the microspheres' surface is relatively smooth (Fig. 2b), which shows more clearly in the high magnification SEM image of a single microsphere in Fig. 2c. The results indicated that fabricated BSA microspheres would support strong optical confinement.

Controlling the size of microlasers is a critical task.[4] For a fixed concentration of BSA solution, the final diameter of dye-doped BSA microspheres depends on their initial droplet size.[25, 29] Fig. 3a shows the dehydration process of four different initial diameter of 100, 94, 89 and 78 μm droplets with the same initial concentration of 500 mg/mL. It can be seen obviously that larger droplet size takes longer time by 350, 310, 245 and 150 seconds to complete dehydration process, respectively. The inset figure of Fig.3a represents the final diameter ($D_f$) of microspheres after dehydration process is linear with the initial diameters ($D_0$) correspondingly, provides a simple and effective solution to control the size of microspheres.  Figure 3b demonstrates more clearly the shrinking of droplet diameter during the dehydration process from the initial diameter of 78 μm to a 44.5 μm solid microspheres at the end of the process (ESI†, Movie 1). The result indicates that if the initial droplets (from a fixed concentration) are uniform then solid microspheres with the same size can be obtained. Uniform droplets can be made by using a microfluidic[28] or solution printing system[30, 31].

The final concentration of BSA in microspheres affects their structure. The diffusion of water leads to the reduction of droplet's size, and therefore, the concentration of BSA in the droplets increases correspondingly. At a specific time, the concentration of BSA can be calculated using the formula: $C_t=C_0D_0{}^3/D_t{}^3$, shown in Fig. 3c. Four microspheres are different initial diameters, the concentration of BSA in the spheres rises gradually from 500 mg/mL to about 1138 - 1150 mg/mL at the end of the dehydration process. This result is similar to previously published work which the concentration of BSA reaches 1147 ± 32 mg/mL.[25]

Studying the effect of the initial concentration of BSA on the dehydration process is also important in order to control the size of microspheres. Fig. 4a describes the change of the size of four droplets at the same initial diameter of 100 μm. Using different the initial concentration of BSA solutions: 500, 200, 125 and 100 mg/mL, four solid-states microspheres are obtained with diameters of 76, 56.5, 48 and 45.5 μm, respectively. The inset figure of Fig. 4a shows the change in the ratio of $D_f$ and $D_0$ of them, which is proportional to $(C_0)^{1/3-}$ fit perfectly with the theory. To obtain smaller microspheres from the same initial size of droplets by reducing the initial concentration of BSA, it takes more time to complete the dehydration process. The longest required time to make ~100 μm-diameter microspheres are about 10 minutes, which is much faster in comparison with the fabrication time of other methods such as the 12 hours by using the vacuum freeze-drying method.[7] Fig. 4b shows the increase in the concentration of BSA in the droplets during the dehydration process. From the initial concentration of BSA solutions: 500, 200, 125 and 100 mg/mL, the same final concentration of BSA of microspheres can reach 1132 - 1138 mg/mL. This means that the initial BSA concentration affects insignificantly to the BSA concentration at the end of the dehydration process. The final concentration of microspheres only depends on the equilibrium of system (i.e. the equilibrium between inner and outer of the droplet). In this case, with the same outer environment as decanol, the dehydration processes with different initial conditions will always stop at the same final concentration of BSA.

Fabricated dye-doped microspheres can act as excellent lasers under optical excitation. Figure 5a shows the schematic of WGM in a typical microsphere. The light is trapped inside by multiple total internal reflections at the microsphere-air interface and subsequently amplified by resonant circulation, leading to laser emission. Figure 5b presents the emission spectra from a single microsphere (47.6 $\mu$m, initial concentration of BSA is 500 mg/mL). It can be seen that the photoluminescence (PL) intensity increases with increasing pump pulse energy (PPE) of the incident laser, and lasing emission is seen when PPE = 0.99 $\mu$J per pulse. The lasing modes are well recognized above the fluorescence background, with sharp peaks appear clearly in the wavelength range from 615 to 645 nm. Besides, the integrated PL intensity is shown in Figure 5c. A nonlinear increase of the emission intensity supports the lasing action, indicating a lasing threshold, which is about 0.75 $\mu$J. The threshold is equivalent to 7.8 $\mu$J/mm$^2$- is comparable with the previous report (same material and shape),[8] and about two orders of magnitude smaller than starch-based biolasers.[7] Lasing modes can be well explained by WGM theory, in particular, by using the explicit asymptotic formula.[32] By assuming the diameter of the sphere as $D$ = 47.595 $\mu$m, the mode number of the transverse electric (TE) and transverse magnetic (TM) modes are calculated to be 333 to 340, which fit very well with the experimental measurements (Fig. 5d).

Size-dependent lasing spectra of WGM lasers were studied and the results are shown in Fig. 6. Figs. 6a-6d plot lasing spectra of four different microspheres. It can be seen that the free spectral range (FSR) decreases with increasing microsphere diameter. It is understandable because the FSR of a WGM laser can be determined as FSR = $\lambda^2/\pi nD$, where $\lambda$ is lasing wavelength, $n$ and $D$ are the refractive index and diameter of the microsphere.[33] The FSR of a 32 $\mu$m-diameter microsphere is determined to 2.5 nm (Fig. 6a). Considering the resonant wavelength $\lambda$ = 623 nm and refractive index $n$ = 1.47, then the calculated FSR is 2.6 nm, which is very close to the experimental observation. When the size of the laser increases, its FSR decreases correspondingly. The FSR of a 123 $\mu$m-diameter microsphere is 0.7 nm (Fig. 6d) which again agrees well with the calculation assuming $\lambda$ = 630 nm. From the above equation, FSR of various microspheres is expected to be linear with the inverse of the microsphere diameters. As shown in Fig. 6e, we experimentally obtained this linear relationship by measuring FSR of 21 microspheres with diameters ranging from 24 to 128 $\mu$m. The result further confirms the WGM mechanism for laser action in the microspheres.

In addition to the lasing spectrum, the Q factor is also expected to be dependent on microsphere size. Figures 7a and 7b show the profile of a typical lasing mode in the range of 1 nm of two different microspheres. It can be seen that the spectral linewidth or the full width at half maximum ($\delta\lambda$) of a 24 $\mu$m-diameter microsphere is 0.36 nm, which nearly doubles compared with that of 0. 2nm of a 123 $\mu$m-diameter microsphere. It means that a larger microsphere exhibits narrower lasing modes compared with a smaller one. In addition, the

Q factor of a lasing mode can be defined as Q = $\lambda/\delta\lambda$. As a result, the Q factor of microspheres should increase with their sizes. Indeed, we obtained a Q factor of about 1700 for a 24 $\mu$m-diameter microsphere and 3100 for a 123 $\mu$m-diameter microsphere. Figure 7c indicates that the Q factor is found to increase with the microsphere diameters. This effect has been previously observed and well explained.[34] In addition, the Q factor of our microsphere biolasers (about 1700 to 3100) is comparable with protein-based microsphere[8] and microdisk[18] lasers using different fabrication methods. In comparison with microlasers using other materials, such as starch-based microlasers,[7] the Q factor of our lasers is about 2 times higher. The lower Q factor of the starch-based laser may be due to its unusual ellipsoid shape (rather than a sphere) which does not effectively confine light.

It is noted that all the lasing properties have been measured in air. Previous reports have demonstrated that BSA-based microlasers can work efficiently in an aqueous environment.[8, 18] However, in this work, we have found that our microlasers tend to swell in water. This issue may be solved by annealing the microlasers at high-temperature (around 100 °C) and we will investigate it in future work.

## 4. Conclusions

We have demonstrated a green and fast process for fabricating solid-state dye-doped BSA microsphere biolasers. The technique is based on the dehydration process via modified Microglassification™ technology. Our approach can produce perfect microspheres within 5-10 minutes - about two orders of magnitude faster when compared with the freeze-drying vacuum method. The sizes of microcavity can be tunable flexibly from 10 to 150 $\mu$m by changing initial BSA concentration or initial diameter of droplets. WGM lasing emission with the lasing threshold of about 8 $\mu$J/mm$^2$ and Q factor of 1700 to 3100 from the dye-doped microspheres. Lasing mechanism and size-dependence of lasing characteristics were investigated, showing an agreement between experimental observation and theoretical calculation. Owing to the fabrication simplicity and fast processing, microspheres biolasers can be mass production, which may allow their use in practical applications including biosensing and bioimaging.

## Acknowledgements
This research is funded by Vietnam National Foundation for Science and Technology Development (NAFOSTED) under grant number 103.03-2017.318.

## Conflicts of interest

There are no conflicts to declare

# ARTICLE

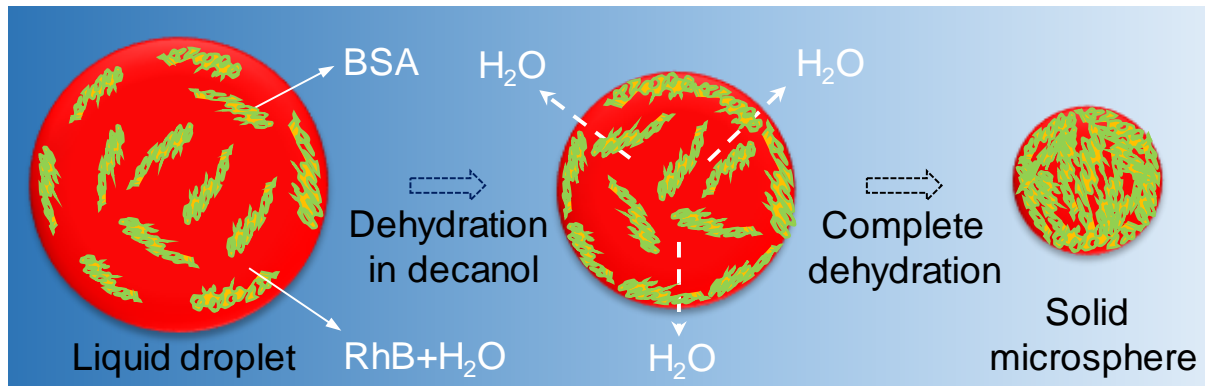

**Fig. 1** Illustration of the dehydration process of a dye-doped liquid droplet in decanol.

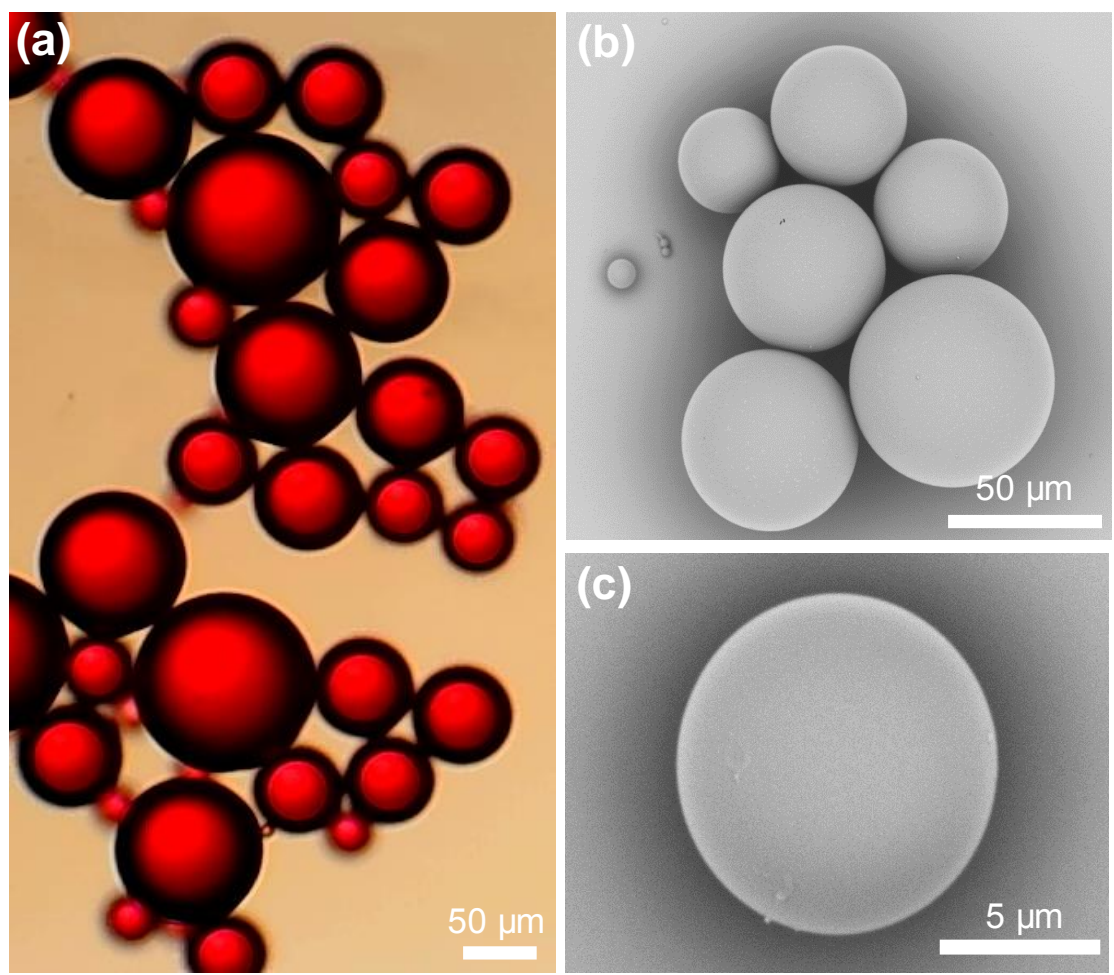

**Fig. 2** (a) Optical microscope image of dye-doped BSA microspheres. (b) Scanning electron microscope (SEM) image of the BSA microspheres with different sizes and (c) High magnification SEM of a single microsphere.

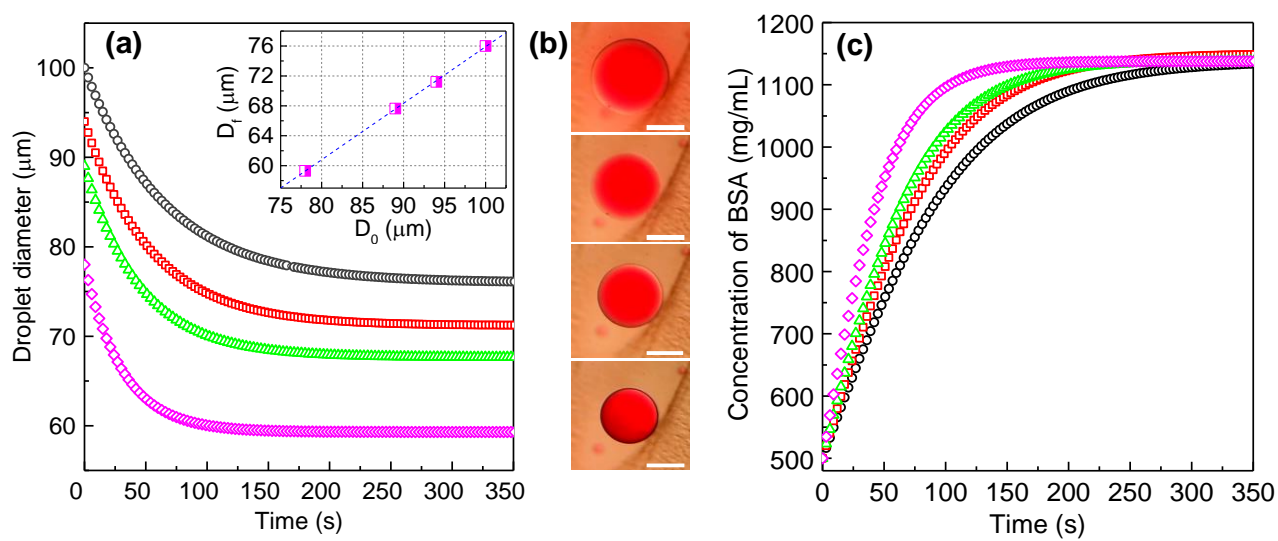

**Fig. 3** (a) Dehydration process of four different dye-doped BSA droplets. The initial concentration of BSA is 500 mg/mL. The inset figure shows the relationship between the final diameter ($D_f$) of microspheres and its initial diameter ($D_0$). (b) Optical microscope images of a 78 μm-diameter BSA droplet during the dehydration process at 0, 40, 80 and 150 seconds, respectively. All scale bars are 50 μm. (c) Change of BSA concentration in the droplets during the dehydration process.

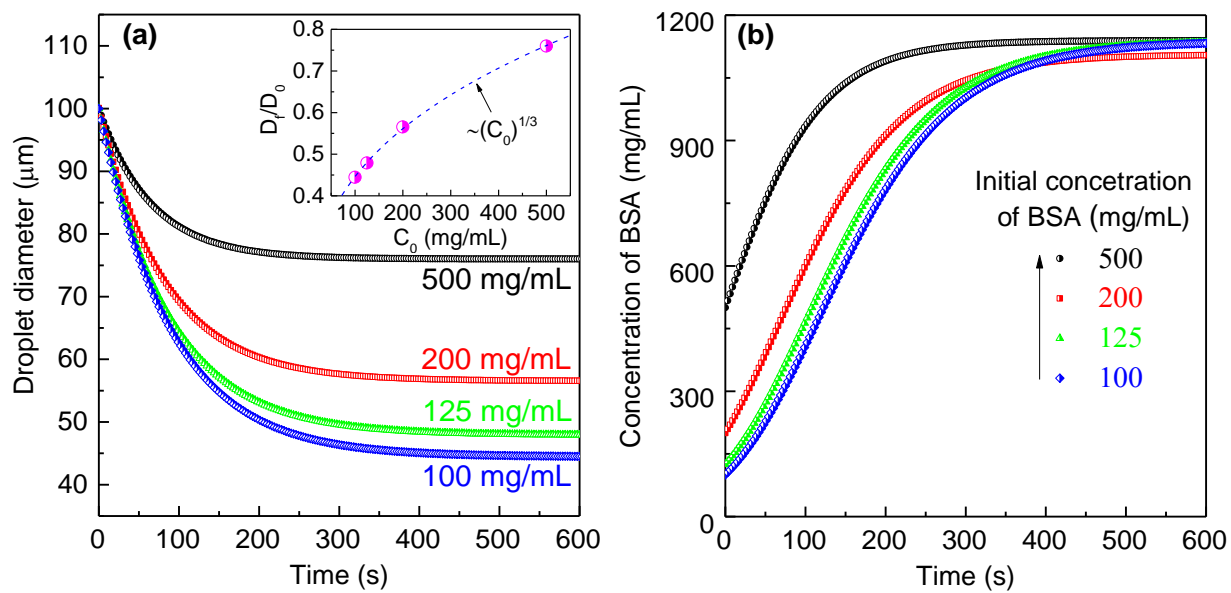

**Fig. 4** (a) Dehydration process of four BSA droplets with different concentrations. The initial diameters of the droplets are similar and approximate of 100 μm. The inset figure shows the ratio of the final diameter and the initial diameter of the droplets ($D_f/D_0$) as a function of the initial BSA concentration. (b) Change of BSA concentration in droplets with different initial concentrations.

**ARTICLE**

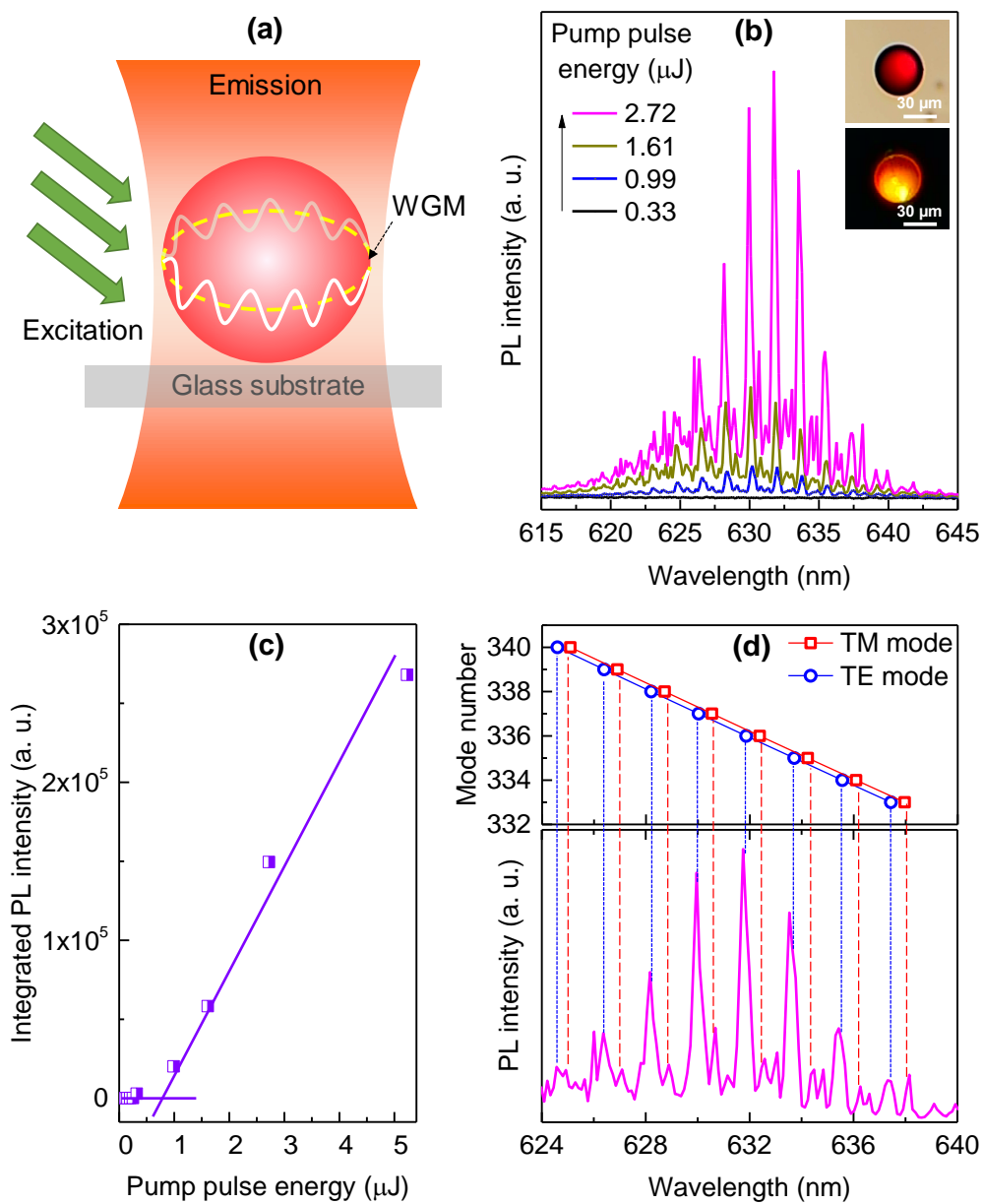

**Fig. 5** (a) Schematic of whispering gallery mode (WGM) in a microsphere. (b) Emission spectra of a 47.6 μm-diameter dye-doped BSA microsphere under pulse excitation. (c) The corresponding integrated PL intensity as a function pump pulse energy. (d) Matching between calculated lasing modes and experimental lasing wavelengths.

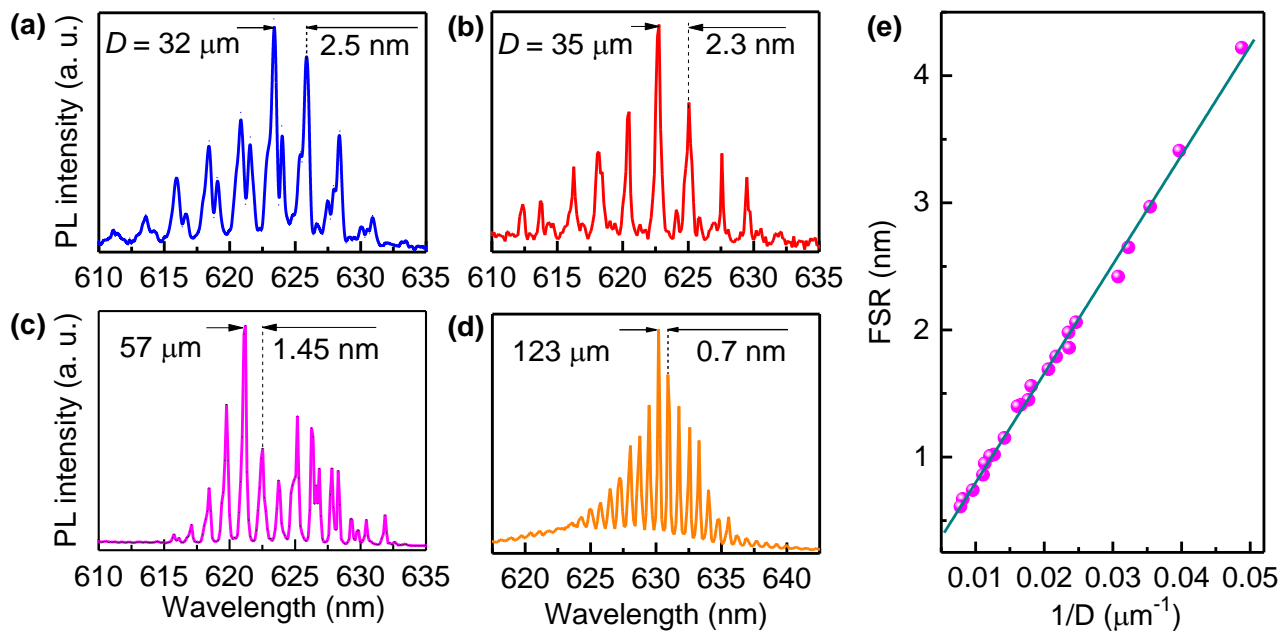

**Fig. 6** (a)-(d) Emission spectra of four different dye-doped BSA microspheres with increasing diameter. (e) Experimental free spectral range (FSR) of microsphere biolasers versus its inverse diameters. The line is theoretical calculation using equation FSR = $\lambda^2/\pi nD$, where $\lambda$ = 625 nm.

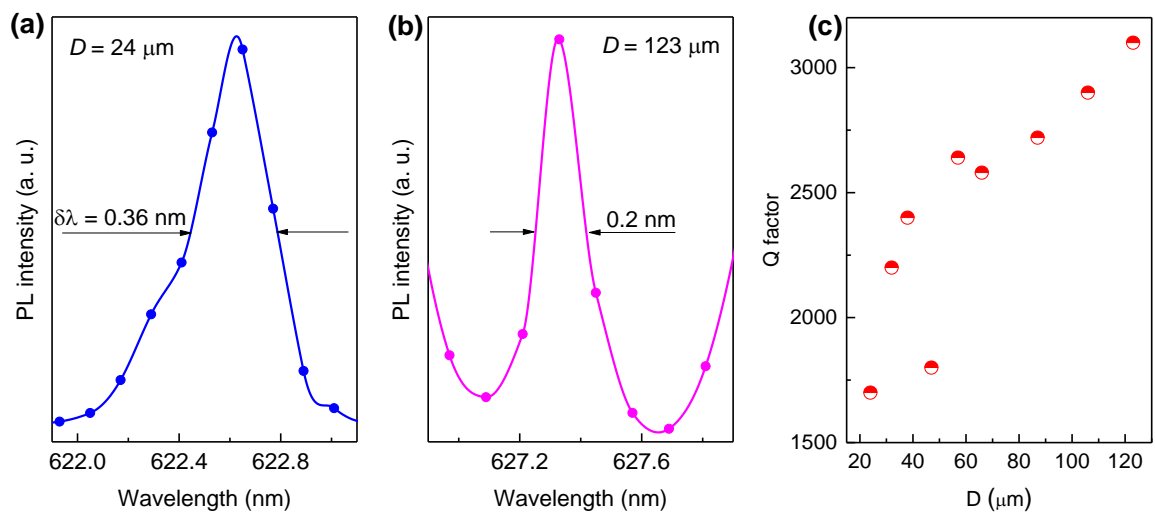

**Fig. 7** (a) and (b) The full width at half maximum (δλ) of a 24 μm-diameter and a 123 μm-diameter microspheres, respectively. (c) Q factor as a function of microsphere diameter.

**Table of Contents Entry**

We demonstrate dehydration as a very fast processing and straightforward method for mass production of high quality protein-based microsphere biolasers.

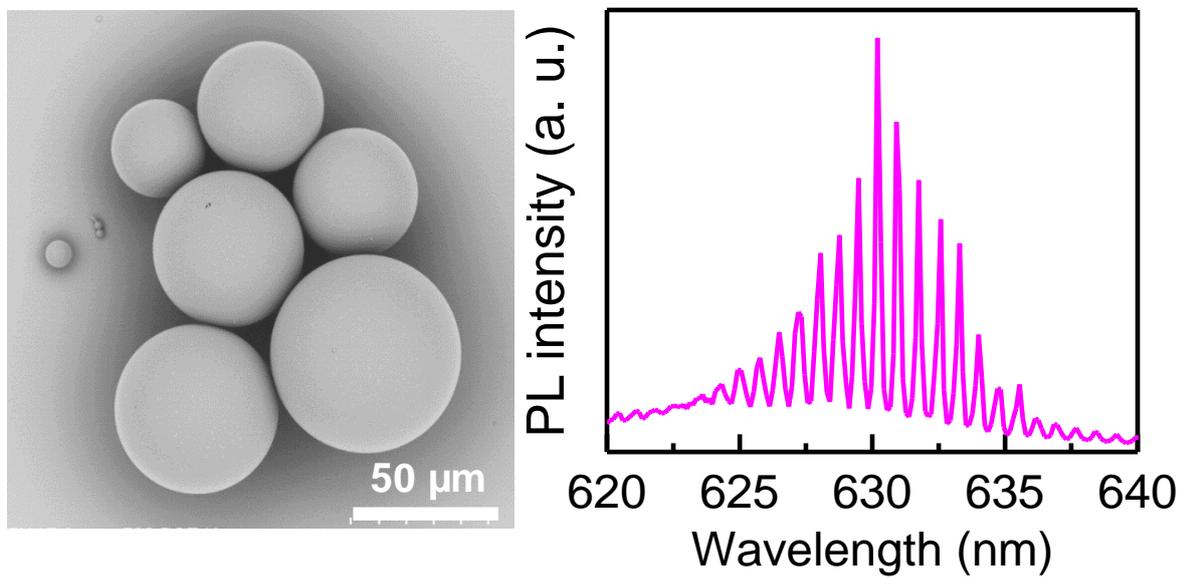